\newcommand\sun{$_\odot$}

\newcommand\arcsec{^{\prime\prime}}
\documentstyle[11pt,aaspp4]{article}
\lefthead{GARNETT ET AL.}
\righthead{BOK GLOBULES IN THE LMC}
\received{7 October 1998}
\revised{6 December 1998}
\accepted{8 December 1998}
\slugcomment{To appear in the March 1999 Astronomical Journal}
\begin{document}
\title{BOK GLOBULES IN THE LMC\altaffilmark{1}}
\author{D. R. Garnett\altaffilmark{2,3}, J. R. Walsh\altaffilmark{4},
Y.-H. Chu\altaffilmark{5}, and B. M. Lasker\altaffilmark{6} }
\altaffiltext{1}{Based on observations with the NASA/ESA Hubble Space
Telescope obtained at Space Telescope Science Institute, which is operated
by the Association of Universities for Research in Astronomy, under NASA
contract NAS5-26555.}
\altaffiltext{2}{Astronomy Department, University of Minnesota, 116 Church St., 
S. E., Minneapolis MN 55455}
\altaffiltext{3}{Current address: Steward Observatory, University of Arizona, 
933 N. Cherry Ave., Tucson AZ 85721; e-mail: dgarnett@as.arizona.edu}
\altaffiltext{4}{Space Telescope European Coordinating Facility, European
Southern Observatory, Karl-Schwarzschild-strasse 2, D-85748
Garching-bei-Munchen, Germany}
\altaffiltext{5}{Department of Astronomy, University of Illinois, 
Urbana, IL 61801}
\altaffiltext{6}{Space Telescope Science Institute, 3700 San Martin
Drive, Baltimore, MD 21218} 
\begin{abstract}
We report the discovery of small, isolated dust clouds in the Large
Magellanic Cloud, which are excellent candidates for 
counterparts to the Bok globules observed in the Galaxy. 
We detect these small clouds silhouetted against diffuse H$\alpha$ 
emission, based on parallel imaging with the WFPC-2 on 
{\it HST}. The clouds we identify as Bok globule candidates have 
typical sizes of approximately 1$\arcsec$, corresponding to 
about 0.25 pc linear diameter at the distance of the LMC. We 
derive lower limits to the optical depth within the dark clouds, 
and masses assuming that the clouds have density distributions 
similar to Galactic Bok globules. The sizes and estimated masses 
for LMC globules are comparable to those estimated for Galactic 
globules. An extended sample of such objects would be 
excellent targets for high-resolution infrared and millimeter 
observations to study low-mass star formation in such clouds 
in low-metallicity environments, where the distance is well known. 

\end{abstract}
\keywords{ISM: globules -- ISM: dust -- galaxies: ISM -- galaxies: 
individual: LMC } 

\section{Introduction}

Fifty years ago, Bok \& Reilly (1947)\markcite{br47} described 
the dark clouds now known as Bok globules, and proposed 
that these clouds were likely undergoing gravitational collapse 
to form new stars. It was many years until that speculation 
was confirmed, however. Since then, a number of recent 
infrared and radio studies have demonstrated that Bok globules 
show a variety of evidence for new stars: embedded warm 
IR sources, often showing evidence for local dust heating, 
nebulosity, or the signatures of protostars (Keene et al. 
1983\markcite{keen93}; Yun \& Clemens 1990\markcite{yc90}, 
1995\markcite{yc90}), Herbig-Haro objects (Reipurth, Heathcote, 
\& Vrba 1992\markcite{rhv92}), and molecular gas outflows (Yun 
\& Clemens 1992\markcite{yc92}). At the same time, detailed 
millimeter-wave emission line studies of some 
of these objects have provided some of the best 
direct evidence yet for {\it infall} of material onto an 
accreting protostar (e.g., Zhou et al. 1993\markcite{zhou93}). 

Isolated Bok globules have thus turned out to be important 
laboratories for the study of protostellar collapse. Nevertheless, 
a few important uncertainties remain regarding the properties of 
the Galactic globules, mainly because distances to individual 
Bok globules are difficult to determine, although new developments
promise to improve this situation in the near future (Peterson
\& Clemens\markcite{pc98} 1998).

In this short article, we present evidence for small, isolated
dark clouds in the Large Magellanic Cloud, discovered 
serendipitously through {\it Hubble Space Telescope} imaging, 
which are likely counterparts to the Galactic Bok globules. 
Although the amount of information we have is limited, we 
describe some derived properties of the LMC globules. Because 
these objects are all at a well-constrained distance, Bok 
globules in the Magellanic Clouds provide an opportunity to 
determine the ensemble properties of globules and their 
embedded protostars, as well as the connection between the 
globules and the local environment.

\section{Observations}

The images discussed here were obtained as part of a parallel 
imaging program of diffuse ionized gas in the Magellanic Clouds; 
a complete description of the program goals and the data obtained 
is given in Walsh et al. (1998, in preparation). All of the 
images were obtained with the WFPC-2 camera on {\it HST}. Of 
the four fields in which globules were found, all four were 
imaged with the F656N filter (H$\alpha$), three were imaged 
in F675W (R band equivalent), three were observed in F547M 
(emission-free V band), and two in F502N ([O~III]). For this 
paper we discuss only the F656N images, which have the highest 
signal/noise. The broad-band images are short exposures intended 
only to identify continuum sources in the F656N band, while the 
[O~III] images are generally less well-exposed than 
the H$\alpha$ images. Total exposure times for the 
H$\alpha$ images ranged from 1800 to 5000 seconds. Table 1 
gives a journal of the H$\alpha$ images and the positions of 
the fields. For a distance of 50 kpc to the LMC (Panagia et 
al. 1991\markcite{pan91}), we can resolve structures as small 
as 0.05 pc (0$\arcsec$.2). 

All of the images were pipeline-processed with the most 
appropriate calibration files. We mosaiced and registered the 
individual exposures for each separate field when necessary, 
then combined them by averaging with the {\it gcombine} task in 
STSDAS, using the `crreject' option to remove radiation hits. 
We generally had only two or three H$\alpha$ exposures per 
field, so CR rejection was not perfect with the long exposure
times involved. We did not apply a correction for charge 
transfer efficiency effects; the dust globules are very 
small-scale objects, and in general H$\alpha$ emission was 
present at an average signal level of at least a few DN all 
across the images, so CTE effects are not expected to exceed 
a few percent across each chip.

\section{Identification and Properties of Globules}

\subsection{Criteria for Identification}

Clemens \& Barvainis (1988\markcite{cb88}; hereafter CB88) 
presented a set of relatively specific criteria for identifying 
a dark cloud as a Bok globule: (1) small size ($<$ 10$^\prime$ 
for Milky Way globules); (2) relative isolation, that is, not 
part of larger dust complexes; (3) no constraint on ellipticity. 
In compiling their catalog of small molecular clouds, they 
rejected dense cores of larger clouds, as well as stringy, 
filamentary dust clouds (which were considered unlikely to be 
hosts for collapsing protostars). 

We employ the CB88 identification criteria to determine what 
dust clouds in our images qualify as candidates for Bok globules. 
Some comments on the identification process are required, however. 
First, there is an obvious scale difference between the Galactic 
objects and LMC candidates. CB88 presented two arguments favoring 
an average distance of about 600 pc for the Galactic Bok globules. 
At a distance of 50 kpc, LMC objects could be expected to be 
smaller by about a factor of 80. The globules in the CB88 catalog 
range in angular size from 1$^{\prime}$ to 10$^{\prime}$, with a 
mean size of 4$^{\prime}$, but with a strong peak at 1-2$^{\prime}$. 
Direct scaling to the LMC distance thus implies angular sizes less 
than 7-10$\arcsec$, with most of the objects around 1$\arcsec$ in 
size. Second is that `isolation' is not always obvious in our LMC 
images: it becomes clear from examination that we are looking at 
multiple overlapping structures, and more than one dust structure 
may be found along a given line of sight. Related to this is the
possibility that the LMC H$\alpha$ emission may have components 
both foreground and background to a dust globule, and in many 
cases a globule may be largely washed out by foreground emission 
line gas. Thus, although we see many structures that may in fact 
be globules, for this paper we point out only the most obvious 
structures. Deep continuum imaging and future millimeter-wave 
interferometer studies may find many more Magellanic Cloud globules.

Within these guidelines, we identify five small dust clouds which 
appear to be excellent candidates to be Bok globules. We list 
these candidates in Table 2 and show an atlas of them in Figure 
1. The coordinates for the rough center of each globule listed 
in the table were obtained using the 'metric' task in STSDAS. The 
sizes that we list are approximately the FWHM along the major and 
minor axes; these are only approximate because the globules do not 
always show symmetric profiles and can be affected by foreground 
H$\alpha$ emission. 

\subsection{Optical Depths and Extinction Estimates}

For these five globules, we provide estimates in Table 2 for the 
peak absorption optical depth in the F656N band, determined from 
the relation

\begin{equation}
{I\over I_0} = e^{-\tau}.
\end{equation}

This estimate is affected by a number of sources of contamination. 
Our H$\alpha$ images in general show emission at some level all
across each field, and we are unable to identify blank regions
to subtract the background, which arises from several sources. 
First, there is the contribution from the foreground Galactic and 
zodiacal light, as well as possible contamination from geocoronal
H$\alpha$. To estimate the contribution from these sources, 
we examined deep FOS spectra of a blank sky field at 2$^h$ 
59$^m$ 47$^s$, $-$20$^{\circ}$ 13$^{\prime}$ 06$\arcsec$, taken 
with the G570H grating (program GO-5968; PI Freedman). The total 
8800 second exposure yielded a surface brightness within the F656N 
bandpass of about 7$\times$10$^{-17}$ ergs cm$^{-2}$ s$^{-1}$ 
arcsec$^{-2}$. This converts to a count rate of about 
1.1$\times$10$^{-3}$ counts per second per pixel in the WFPC-2 
images. In comparison, the WFPC-2 Exposure Time Calculator predicts 
a background count rate of 9$\times$10$^{-4}$ counts per second 
per pixel at the blank sky position, and 7$\times$10$^{-4}$ at 
our LMC positions. For our purposes, we use a background count rate 
of 10$^{-3}$ counts per second per pixel, with an uncertainty of 
$\pm$50\%. This contributed less than 10\% to the observed H$\alpha$ 
signal. We scaled the countrates by the effective exposure 
times of our averaged images and subtracted the background before
computing the optical depth. A second source of contamination is 
foreground H$\alpha$ within the LMC. It is apparent in some cases 
that there is H$\alpha$ emission in front of the dust globules, 
which will cause us to underestimate the optical depth. We make 
no correction for such emission but note that it is likely to be 
present, and we treat the optical depths as lower limits only.

Our lower limits for $\tau$(6563) correspond to lower limits on 
the visual extinction $A_V$ of 1.5-3 magnitudes, based on the 
Galactic extinction curve for $R_V$ = 3.1 (Cardelli, Clayton, 
\& Mathis 1989\markcite{ccm89}); at the wavelengths considered,
the Galactic extinction curve is sufficiently similar to the LMC
curve, especially for small values of extinction. On the other 
hand, the nature of dust grains in metal-poor environments such
as the LMC is not well known. This is an area greatly needing
future investigation.

As expected, our derived values for $A_V$ are significantly lower 
than those obtained for Galactic globules from star counts. It is 
likely that diffuse foreground H$\alpha$ emission, which is seen 
all across our F656N images, contributes significant emission in 
the cores of the globules. A deep broad-band continuum study would
be better suited for estimating accurate extinctions within 
these clouds.

\subsection{Estimates of Globule Masses}

We can estimate roughly the masses of the LMC globules by 
extrapolating the properties of Galactic globules. Clemens, Yun,
\& Heyer\markcite{cyh91} (1991) inferred an average H$_2$ number 
density $<$N(H$_2$)$>$ $\approx$ 10$^3$ cm$^{-3}$ from millimeter 
studies of CO in Galactic Bok globules. If we approximate the LMC 
clouds as spheres with a radius $r$ = $(ab)^{0.5}$, (where $a$ and 
$b$ are the semi-major and semi-minor axes as obtained from the
sizes listed in Table 2), and if 
we assume the LMC clouds have the same average H$_2$ density as 
Galactic globules, we obtain the cloud masses listed in Table 2. 
The masses range between 0.3 and 80 solar masses, with four of 
the five clouds under one solar mass. In comparison, Clemens et 
al. (1991) found a range of 0.6 to 200 solar masses for the 
Galactic sample, with an average mass of about 11 M$_{\sun}$. It 
is possible we may underestimate the LMC globule masses somewhat 
if foreground H$\alpha$ emission causes us to underestimate the 
angular sizes of the clouds. For example, underestimating the 
sizes of the smallest globules by only 25\% (abut 2 WFC pixels) 
would lead us to underestimate their masses by a factor of two. 
Again, deep broad-band imaging could reduce the uncertainty.

The discrepancy between the typical masses for our LMC globules
and the Milky Way globules is curious, and may bring the assumed
density into question. If we use our estimated extinctions and
the diffuse gas N(H)/E(B$-$V) ratio for the LMC 
(Fitzpatrick\markcite{fitz85} 1985, Koorneef\markcite{koor82} 
1982), we obtain densities and masses about an order of magnitude 
larger than those listed in Table 1. On the other hand, it is
likely that our estimated extinctions are severe underestimates,
and it is not clear that the diffuse gas dust/gas ratio should
be applicable to dark clouds. Note that the four small clouds in 
our LMC sample would be among the smallest objects in the CB88
sample and thus might be expected to have small masses as well. 
We also note that the recent distance estimate for the Milky Way 
globule CB24 (Peterson \& Clemens\markcite{pc98} 1998) results
in a mass estimate only one-fifth of that based on the estimated
average distance of 600 pc. It may be that many galactic globules
have quite low masses as well. The mass function of LMC globules
will become better defined as more globules are detected in high
resolution images.

\subsection{Space Density of LMC Globules}

We can estimate crudely the number of small globules that might 
exist in the LMC from the surface density of detected globules. 
We detected five small globules in four WFPC2 fields, which cover 
an angular area of 23 square arcminutes. At a distance of 50 kpc, 
this translates to a globule surface density of approximately 
1000 per kpc$^2$ (not corrected for the LMC's inclination). If we 
assume that the small globules are distributed over the full 50 
square degrees of the main body of the LMC, we would predict a 
total of 4$\times$10$^4$ LMC globules. For comparison, Clemens et 
al. (1991) estimated a total of 3.2$\times$10$^5$ small globules 
in the Milky Way, corresponding to a surface density of about 
400 globules per kpc$^2$ within 15 kpc of the galactic center. 
Thus, the simplest calculation leads us to estimate that 
the LMC is forming globules at about twice the rate per unit 
area as the Galaxy. By coincidence, this is also the ratio 
of the estimated star formation rates per unit area for 
the two galaxies (R. C. Kennicutt, private communication).

One could easily question some of the assumptions behind this
simple estimate. First, we have already noted that our sample 
is biased toward the most obvious globules. We could be missing 
many objects because of low contrast due to the strong foreground 
emission in this area. On the other hand, there is no evidence 
that globules should be distributed homogeneously across the LMC 
main body. It may be that such structures are largely located 
in the region of dark clouds/molecular gas and strong star 
formation in the vicinity of 
30 Doradus, where our fields are located. Hodge\markcite{hod88} 
(1988) cataloged 146 dark clouds in the LMC based on CTIO 
4m prime focus photographic plates. However, 115 of these clouds 
were located in two fields at the east end of the 
LMC bar, near 30 Dor. If we assume that small globules are 
preferentially located in these fields (covering a little over 
one square degree), we would predict about 1000 globules would 
be found in the LMC. Therefore, although the coincidence between 
the ratios of globule surface densities and star 
formation rates is intriguing, the numbers are too uncertain at 
present to interpret reliably. As more fields are imaged, the 
estimated number of globules in the LMC will improve greatly.

\section{Other Dust Structures}

Figures 2 and 3 show a number of other structures which are 
analogous to structures observed in Galactic dust clouds. 
Figure 2 shows an ``elephant trunk'' structure in the same 
field as globule BGJ053933-691338. This feature is about 
12$\arcsec$ (about 3 pc) long, but only about 
0.1 pc wide in its narrowest parts. Such coherent dust 
structures are thought to be clouds stretched out by 
the interaction between the dark clouds and a less 
dense flow of neutral or ionized gas past the cloud (e.g., 
Schneps, Ho, \& Barrett\markcite{shb80} 1980), although 
magnetic fields may play an important role 
as well (Carlqvist, Kristen, \& Gahm\markcite{ckg98} 1998).

Figure 3 shows a larger dust cloud, in the same field as 
BGJ053615-691822, which shows evidence for several dense clumps, 
similar in size to the isolated globules. 
This may be a molecular cloud with multiple cores 
that will eventually become a cluster of stars. 
It is interesting that the densest clumps appear to be 
on a side of the dark cloud that is sharply defined by a 
photoionized edge. This may be a case in which gravitational 
collapse is organized and possibly triggered by the mechanical 
and radiation energy from the hot massive stars that are 
responsible for the ionized gas. A survey for IR and millimeter 
sources in this region could prove interesting. It is also 
interesting to note that to the lower left of this cloud 
in Figure 3 is a star which is immediately adjacent to a 
small, arc-shaped dust cloud. This star could be a very 
young star that has just broken out of its dusty cocoon. 

\section{Discussion}

It is very likely that the small globules described in this
paper are the smallest dust structures ever detected in any
galaxy other than the Milky Way. The smallest dust structures 
detected by Hodge (1988) in the LMC were about 2 pc across, 
about 10$\arcsec$ at the distance of the LMC. The high-resolution 
imaging capability of $HST$ has allowed us to push this 
scale down another factor of ten. At this spatial scale the 
complexity of the structure for both emission and dust absorption 
in the LMC becomes very evident. Hodge (1988) noted the 
difficulty of detecting dark clouds against a faint and 
variable background. This is also true of our WFPC images, 
to which we add the confusion caused by overlapping structures 
and the low contrast due to foreground emission. Hodge 
also noted a curious lack of correspondence between CO emission 
(Cohen et al.\markcite{cohen88} 1988) and dark clouds in 
some parts of the LMC. This may reflect partly the low resolution 
of the CO surveys, but it also suggests that both visual 
surveys for dark clouds and CO emission studies are necessary 
to characterize the population of dark clouds in galaxies.

In many ways the parallel imaging shows that the structure 
of the ISM and dust clouds in the LMC is very similar to that 
observed in the Galaxy. We see in our LMC fields many of the
same dust structures that have been identified in the Galaxy, 
and the tentative inferred properties of those structures 
appear to be similar to those of their Galactic counterparts,
although the estimated masses appear to be a bit low. 


Future studies of globules in the Magellanic Clouds are of
importance for understanding star formation in very different
environments from the Milky Way. The Clouds are at comparatively 
well-determined distances, and so the ensemble properties of 
LMC and SMC globules in principle could be well constrained 
with high-resolution infrared and millimeter studies. Potentially 
interesting future observations include high resolution K-band 
imaging surveys for protostars within the globules. The 
near-IR sources discovered by Yun \& Clemens (1994)\markcite{yc94} 
in Galactic Bok globules have K-band magnitudes between 6 and 13. 
Assuming an average distance of 600 pc for the Galactic objects, 
the same near-IR sources would have K = 16-23 in 
the Magellanic Clouds. Deep IR imaging of LMC and SMC globules, 
especially with future 8-m 
class telescopes or {\it HST}, will likely detect embedded 
protostellar sources if they exist. Such observations 
will begin to constrain the properties of low-mass protostars 
in the Magellanic Clouds. 

High resolution ($\approx$ 1$\arcsec$ or better) molecular line 
observations would also be of great interest for density and 
mass determinations, from which the mass function of globules 
could be derived. This would help provide a refined estimate 
of the star formation efficiency in globules, and may facilitate 
comparisons between isolated globules and the embedded dark 
cores in larger clouds. Also of interest is whether the 
lower dust-to-gas ratios in the Magellanic Clouds systematically 
affects the properties of globules compared to their 
Galactic counterparts. The lower dust-to-gas ratio leads to 
greater penetration of UV radiation into the dark clouds, 
affecting the cloud structure. This is one suggested cause for 
the lower I(CO)/N(H$_2$) ratios observed in the Magellanic Clouds 
and other dwarf irregulars (Poglitsch et al.\markcite{pog95}
1995; Israel et al.\markcite{isr96} 1996; Madden et 
al.\markcite{mad97} 1997), and may lead to evaporation of small 
dark clouds. Measurements of velocity widths should show 
whether the LMC globules are virialized and stable or expanding 
and evaporating. Finally, the largely unobscured view 
of the structure of the Magellanic Clouds allows one to examine 
the isolated globules in the context of large-scale star 
formation and the relation to other dark cloud structures. 

\acknowledgments
We thank Tony Keyes at STScI for assistance in locating deep 
blank-field FOS spectra to estimate the H$\alpha$ foreground. 
DRG thanks Neal Evans for helpful discussions regarding the
potential interest of Bok globules in the Magellanic Clouds. 
We are grateful to the anonymous referee for several excellent
suggestions which enhanced the content of this paper. Support 
for the U.S. investigators on the 
parallel imaging program was provided by NASA and 
STScI through grants GO-3589-91A, and GO-4497-92A. DRG acknowledges 
support from NASA-LTSARP grant NAG5-6416, as well as 
the sponsorship of the European Southern Observatory in April/May 
1998, where much of this paper was written. YHC and 
DRG also acknowledge the hospitality of the Aspen Center for 
Physics in June 1998, which provided a very pleasant forum 
for discussions of the results presented here.


\clearpage

\begin{figure}
\plotone{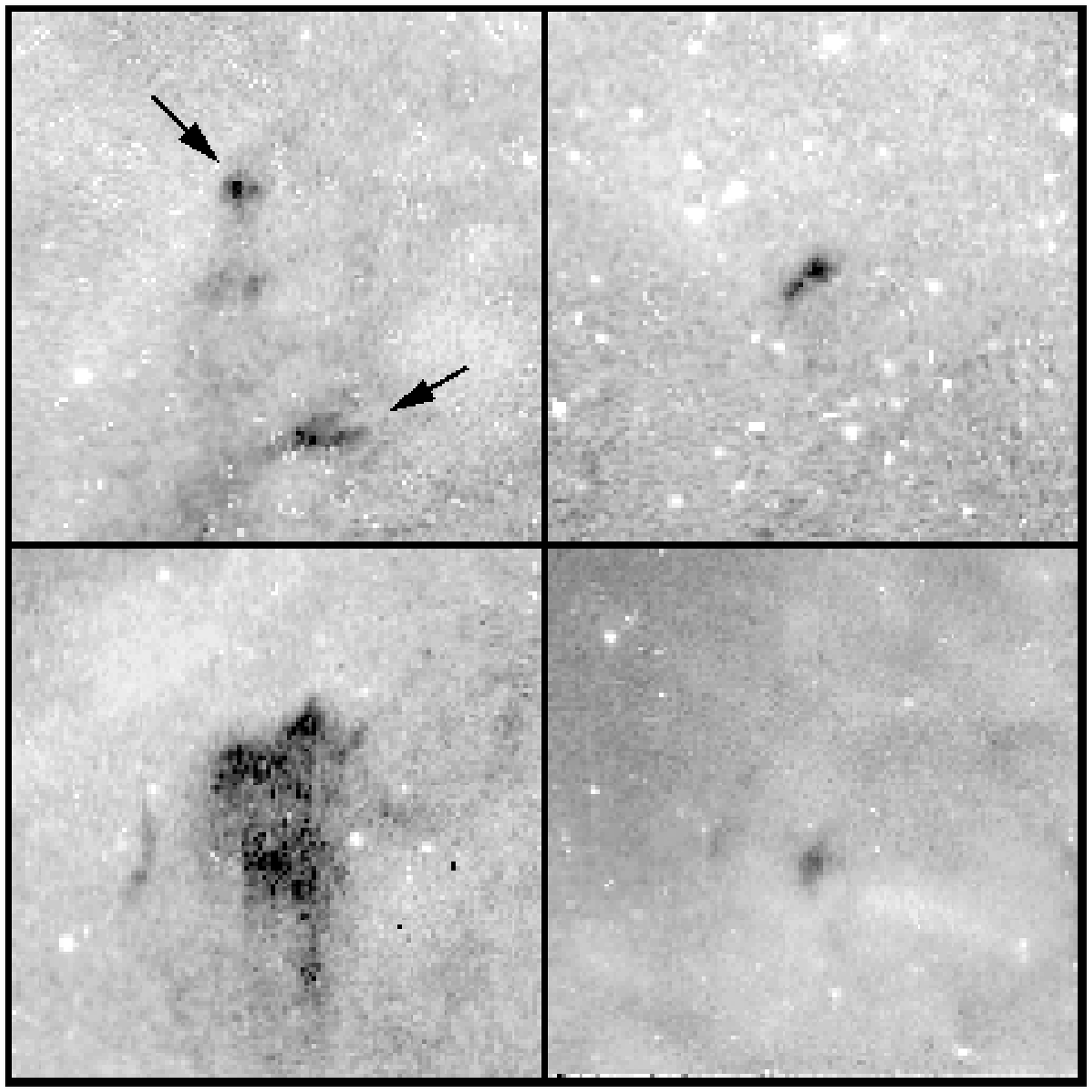}
\figcaption[Garnett.fig1.ps]{WFPC2 F656N images of candidate Bok 
globules in the LMC. Each frame is 15$\arcsec$ on a side. The images
are displayed on a log scale with the same range of values. {\it 
Upper left:} LMC BGJ053703-690616 (upper) and BGJ053702-690610 
(lower). {\it Upper right:} LMC BGJ053805-691813. {Lower left:} 
LMC BGJ053933-691338. Note that there is significant foreground 
H$\alpha$ emission in front of this cloud. {Lower right:} LMC 
BGJ053615-691822.  }
\end{figure}


\begin{figure}
\plotone{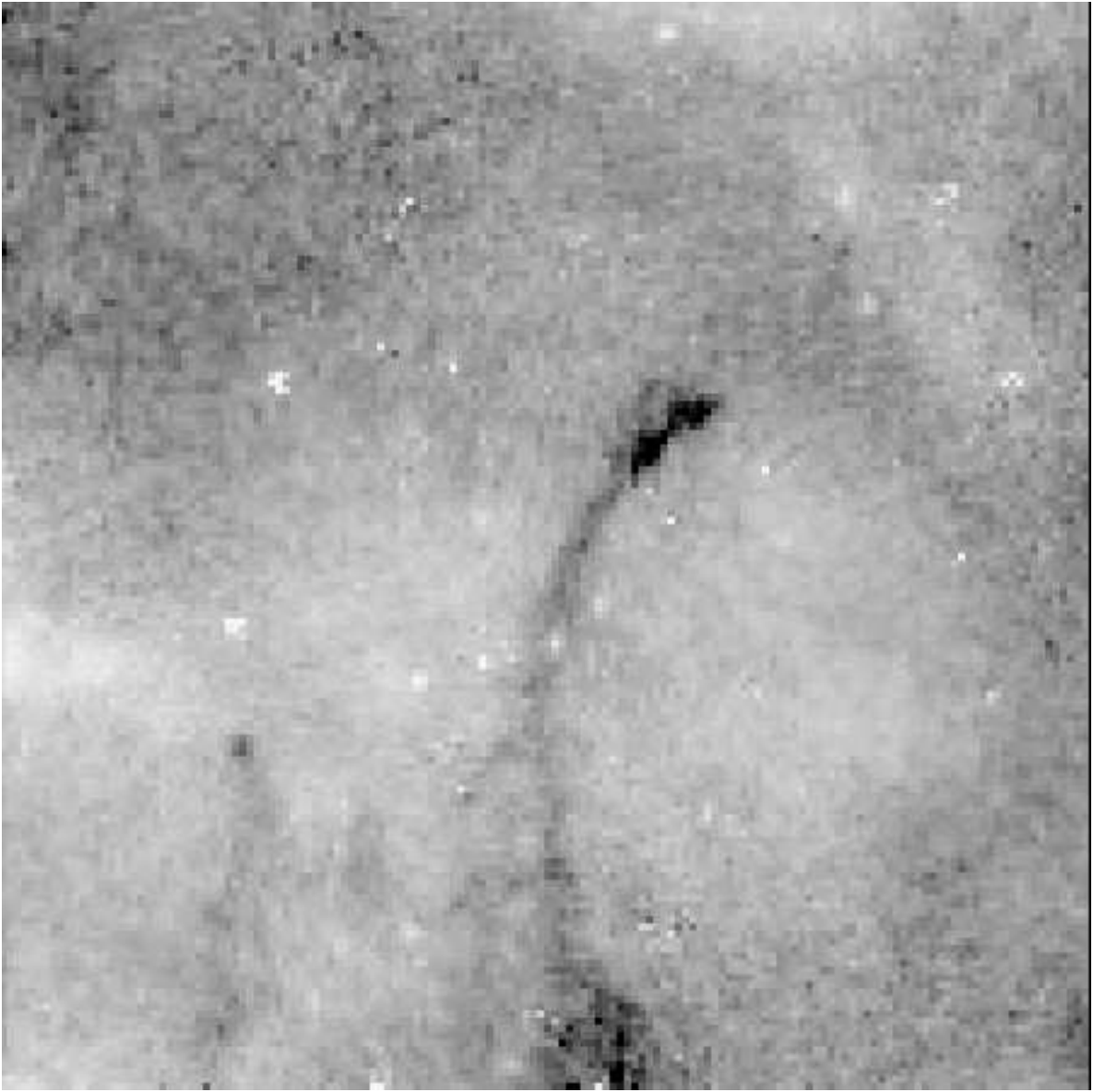}
\figcaption[Garnett.fig2.ps]{Portion of the WFPC2 F656N image of 
field 3, showing elephant trunk structures. The frame is 20$\arcsec$
on a side. }
\end{figure}


\begin{figure}
\plotone{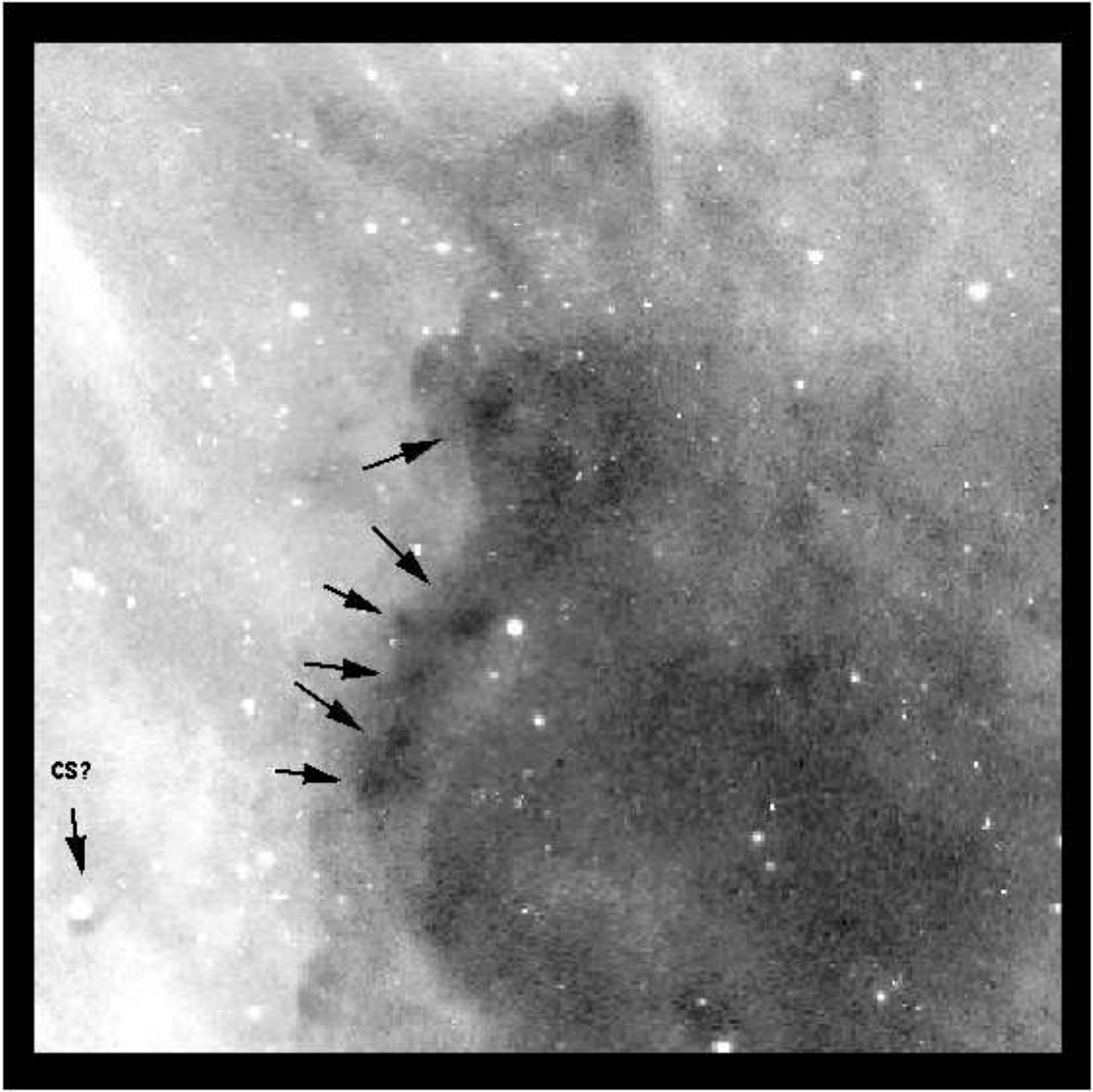}
\figcaption[Garnett.fig3.ps]{Portion of the WFPC2 F656N image of field 
4, showing a dark cloud that appears to have multiple cores (arrowed). 
Note also the star in the lower left (labeled CS?) which is on the edge 
of a small dark cloud. The image scale is 40$\arcsec$ on each side. 
The contrast appears low because of foreground H$\alpha$ emission.}
\end{figure}

\clearpage

\begin{deluxetable}{lccc}
\tablenum{1}
\tablewidth{35pc}
\tablecaption{Journal of WFPC-2 F656N Exposures }

\tablehead{
\colhead{Observation ID} & \colhead{RA (J2000)} & \colhead{Dec (J2000)} & \colhead{Exposure Time} \nl
}
\startdata
field 1:  &            &                &         \nl
U29J1W01T & 5:37:00.02 & $-$69:07:21.28 &  900s  \nl
U29J1X01T & 5:37:00.02 & $-$69:07:21.28 &  900s  \nl
field 2:  &            &                &         \nl
U2OR1V01T & 5:37:55.79 & $-$69:17:18.35 & 2200s  \nl
U2OR1Z01T & 5:37:55.92 & $-$69:17:19.97 & 1600s  \nl
field 3:  &            &                &         \nl
U3820Q01T & 5:39:35.88 & $-$69:12:40.56 & 1700s  \nl
U3820R01T & 5:39:35.53 & $-$69:12:39.97 & 1400s  \nl
U3820S01T & 5:39:35.53 & $-$69:12:34.56 & 1100s  \nl
field 4:  &            &                &         \nl
U3820Z01R & 5:36:12.02 & $-$69:19:36.86 & 2200s  \nl
U3821101R & 5:36:12.02 & $-$69:19:36.86 & 1400s  \nl
U3821102R & 5:36:12.02 & $-$69:19:36.86 & 1400s  \nl
         &            &                &         \nl
\enddata
\end{deluxetable}

\clearpage

\begin{deluxetable}{lccccc}
\tablenum{2}
\tablewidth{44pc}
\tablecaption{Derived Properties of LMC Bok Globules }
\tablehead{
\colhead{Object} & \colhead{Coordinates } & \colhead{Ang. Size } & 
\colhead{Linear Size} & \colhead{Peak } & \colhead{Est. Mass} \nl
\colhead{} & \colhead{(J2000)} & \colhead{(arcsec)} &
\colhead{(parsecs)} & \colhead{$\tau$(6563)} & \colhead{(M$_{\sun}$)} 
}
\startdata
BGJ053703-690616 & 5:37:02.66 ~ $-$69:06:16.2 & 0.7x0.9 & 0.2x0.2 & $>$ 1.2 & 0.3 \nl
BGJ053702-690610 & 5:37:02.38 ~ $-$69:06:10.3 & 0.8x1.6 & 0.2x0.4 & $>$ 1.1 & 0.8 
\nl
BGJ053805-691813 & 5:38:05.13 ~ $-$69:18:13.0 & 0.7x0.9 & 0.2x0.2 & $>$ 1.5 & 0.3 
\nl
BGJ053933-691338 & 5:39:32.76 ~ $-$69:13:37.7 & 4.2x7.1 & 1.0x1.8 & $>$ 2.4 & 80 
\nl
BGJ053615-691822 & 5:36:14.94 ~ $-$69:18:22.3 & 0.6x1.2 & 0.2x0.3 & $>$ 2.0 & 0.5 
\nl
\enddata
\end{deluxetable}

\end{document}